\documentclass[preprint,showpacs,superscriptaddress,prd]{revtex4}
\usepackage{graphicx,amsmath,dcolumn,bm}
\begin{document}
\title{\Large{\bf{Energy loss of charm quarks from  $J/\psi$ production in cold nuclear matter  }}}
\author{Li-Hua Song }
\email[E-mail: ]{songlh@mail.heuu.edu.cn}
\affiliation{College of
Science,  Hebei United University, Tangshan 063009,
P.R.China}
\affiliation{Department of Physics, Hebei Normal
             University,
             Shijiazhuang 050024, P.R.China}



\author{Wen-Dan  Miao }
\email[E-mail: ]{miaowd@mail.hebtu.edu.cn}

\affiliation{Department of Physics, Hebei Normal
             University,
             Shijiazhuang 050024, P.R.China}
\affiliation{Hebei Advanced Thin Films Laboratory, Shijiazhuang
               050024, P.R.China}

\author{Chun-Gui  Duan  }
\email[E-mail: ]{duancg@mail.hebtu.edu.cn}

\affiliation{Department of Physics, Hebei Normal
             University,
             Shijiazhuang 050024, P.R.China}
\affiliation{Hebei Advanced Thin Films Laboratory, Shijiazhuang
               050024, P.R.China}


\begin{abstract}
$J/\psi$ suppression in p-A collisions is studied by considering the
nuclear effects on parton distribution, energy loss of beam proton
and the finial state energy loss of color octet $c\overline{c}$. The
leading-order computations for $J/\psi$ production cross-section
ratios $R_{W/Be}(x_{F})$ are presented and compared with the
selected E866 experimental data with the $c\overline{c}$ remaining
colored on its entire path in the medium. It is shown that the
combination of the different nuclear effects accounts quite well for
the observed $J/\psi$ suppression in the experimental data. It is
found that the $J/\psi$ suppression on $R_{W/Be}(x_{F})$ from the
initial state nuclear effects is more important than that induced by
the energy loss of color octet $c\overline{c}$ in the large $x_F$
region. Whether the $c\overline{c}$ pair energy loss is linear or
quadratic with the path length is not determined. The obtained
$c\overline{c}$ pair energy loss per unit path length
$\alpha=2.78\pm0.81$ GeV/fm, which indicates that the heavy quark in
cold nuclear matter can lose more energy compared to the outgoing
light quark.

\vskip 1.0cm

\noindent{\bf Keywords:} $J/\psi$ production, charm quark, energy
loss.

\end{abstract}

\pacs{ 24.85.+p ; 
       25.40.-h  
       12.38.-t; 
       13.85.-t 
       }

\maketitle
\newpage
\vskip 0.5cm

\section{Introduction}
 $J/\psi$ suppression observed in heavy-ion collisions at RHIC
and LHC, is considered as a most reliable signature for the
formation of Quark-Gluon Plasma (QGP) in the present time. In order to
quantify the properties of the QGP created in heavy-ion collisions,
a solid understanding of the basic mechanisms responsible for the
suppression of $J/\psi$ production due to the nuclear modification
of particle spectra in cold nuclear matter is required. The study
about the nuclear effects on $J/\psi$ suppression in p-A collisions
can give a good baseline for clarifying the conventional nuclear
suppression mechanism in heavy-ion collisions.

The nuclear dependence of $J/\psi$ production cross sections as a
function of the Feynman variable $x_{F}$ has been studied by several
proton-induced fixed target experiments, such as NA3[1], E772[2],
E866[3,4], NA50[5] and HEAR-B[6]. The E866
collaboration[3,4]published the precise measurement of the $J/\psi$
suppression for 800 GeV protons incident on iron and tungsten
nuclear targets, relative to beryllium nuclear targets with very
broad coverage $-0.1< x_{F}<0.95$. The observed suppression is
smallest for $x_{F}\leq 0.25$, and increases at larger values of
$x_{F}$.

$J/\psi$ production is generally believed to occur in two different
steps where a charm-quark pair is produced first through the
interaction of a projectile on a target parton, then followed by the
non-perturbative formation of colorless asymptotic state.  As for
the non-perturbative formation of $J/\psi$,  some approaches
attribute  to an effective absorption cross section $\sigma_{abs}$
of the $c\overline{c}$ pair[7,8]. In fact, if the distance $L_{0}$
which the $c\overline{c}$ pair travels during the proper color
neutralization time $\tau_{0}$  exceeds the length $L_{A}$ of its
path in the medium, the $c\overline{c}$ will remain colored on its
entire path in the medium and the $J/\psi$ hadronization will occur
outside the nucleus. In this case, the so-called nuclear absorption
$\sigma_{abs}$ becomes irrelevant and the energy loss of color octet
$c\overline{c}$ due to the medium-induced gluon radiation is the
main final state effect. Otherwise, the $J/\psi$ hadronization will
occur inside the nucleus and nuclear suppression partly arises from
the absorption cross section $\sigma_{abs}$ of the charm-quark pair.
If the color octet $c\overline{c}$ is produced at a random point
inside a nucleus of radius $R_{A}$, then $L_{A}=3R_{A}/4$. With a
proper colour neutralization time $\tau_{0}=0.25$fm, the relation
$L_{0}\geq L_{A}$ for $x_{F}\geq 0.2$ at $\sqrt{s}=20$GeV and for
$x_{F}\geq 0$ at $\sqrt{s}=40$GeV[9], which implies that the
$c\overline{c}$ will remain colored on its entire path for the E866
experimental data in the kinematical range $x_{F}\geq 0$. Therefore,
$J/\psi$ production in p-A collisions seems the ideal tool for
exploring the energy loss of charm quarks in cold nuclear matter.

In our preceding article[10], by combining three representative sets
of nuclear parton distributions with the energy loss effect of the
incident proton determined by the nuclear Drell-Yan reaction[11], a
leading order phenomenological analysis is performed on $J/\psi$
production cross section ratios $R_{W/Be}(x_{F})$ for the E866
experimental data. The nuclear absorption effect in final state are
taken into account in the Glauber model framework. It is shown that
the energy loss effect with resulting in the suppression on
$R_{W/Be}(x_{F})$ is more important  than the nuclear effects on
parton distributions in high $x_{F}$ region. It is found that the
$J/\psi$-nucleon inelastic cross section $\sigma^{J/\psi}_{abs}$
depends on the kinematical variable $x_{F}$, and increases as
$x_{F}$ in the region $x_{F}>0.2 $. In view of the different idea
for the $c\overline{c}$ pair  color neutralization, the present work
will focus on the final state effects owing to the energy loss of
color octet $c\overline{c}$.  By means of the phenomenological
analysis at the leading order for  $J/\psi$ production cross-section
ratios for the E866 experimental data with the kinematical domain
for a compact color octet $c\overline{c}$ propagating through the
nuclear target, the energy loss effect of charm-quark pair is
researched. It is hoped that our study gives some hint of the energy
loss effects on  the suppression of $J/\psi$ production in heavy-ion
collisions.

The remainder of the paper is organized as follows. In Section II, a
brief formalism for the differential cross section in $J/\psi$
production is given. Section III is devoted to the  results and
discussion. Finally, a summary is presented.

\section{The formalism for $J/\psi$ production differential cross sections }
In the color evaporation model (CEM), quarkonium production is
treated identically to open heavy-quark production except that the
invariant mass of the heavy quark pair is restricted to be less than
twice the mass of the lightest meson that can be formed with one
heavy constituent quark. For charmonium the upper limit on the
$c\overline{c}$ pair mass is then two times the mass of  D meson
mass $m_{D}$. The hadroproduction of heavy quark at leading order
(LO) in perturbative QCD is the sum of contributions from
$q\overline{q}$ annihilation and $gg$ fusion. The charmonium
production cross section ${d\sigma}_{pp}/{dx_{F}}$ is a convolution
of the $q\overline{q}$ and $gg$ partonic cross sections with the
parton densities in the incident proton  and the target proton [12]:

\begin{eqnarray}
\frac{{d\sigma}_{pp}}{{dx_{F}}}(x_{F})&=&\rho_{J/\psi}\int^{2m_{D}}_{2m_{c}}dm\frac{2m}{\sqrt{x_{F}^{2}s+4m^{2}}}
\times[f_{g}(x_{1},m^{2})f'_{g}(x_{2},m^{2})\sigma_{gg}(m^{2})\nonumber \\
& & +\sum^{}_{q=u,d,s}\{f_{q}(x_{1},m^{2})f'_{\bar{q}}(x_{2},m^{2})
+{f_{\bar{q}}(x_{1},m^{2})f'_{q}(x_{2},m^{2})}
\}\sigma_{q\bar{q}}(m^{2})],
\end{eqnarray}
where $x_{1(2)}$ is the projectile proton (target) parton momentum
fractions, $x_{F}=x_{1}-x_{2}$, $\sqrt{s}$ is the center of mass
energy of the hadronic collision, $m^{2}=x_{1}x_{2}s$, $m_{c}=1.2$
GeV and $m_{D}=1.87$ GeV are respectively the charm quark and D
meson mass, and $\sigma_{gg}(\sigma_{q\bar{q}})$ is the LO
$c\overline{c}$ partonic production cross section from the gluon
fusion (quark-antiquark annihilation). $\rho_{J/\psi}$ is the
fraction of $c\overline{c}$ pair which produces the $J/\psi$ state,
$f_{i}$ and $f'_{i}$ stand respectively for the parton distribution
function in the incident proton and in the target proton.

As for p-A collisions, if only considering the nuclear effects of
parton distribution functions, $J/\psi$ differential production
cross section ${d\sigma}_{pA}/{dx_{F}}$ is given by replacing parton
distribution functions of the target proton with the bound nucleon
parton distribution functions in equation (1).

With regard to the energy loss of the projectile proton in the
nuclear target, it is supposed that after the projectile proton has
$n$ collisions with nucleons in nucleus, the center of mass system
energy $\sqrt{s'}$ of the nucleon-nucleon collision producing
$c\overline{c}$ is reduced to
\begin{equation}
\sqrt{s'}=\sqrt{s}-(n-1)\Delta\sqrt{s},
\end{equation}
where $\Delta\sqrt{s}$ is the center of mass system energy loss (see
Ref.[10,11]for more detail discussion). The $J/\psi$ production
cross section in the nth collision can be rewritten as
\begin{equation}
\frac{d\sigma^{(n)}_{pA}}{d x_{_{\rm
F}}}{(x_F)}=\frac{d\sigma_{pA}}{d x^{\prime}_{_{\rm
F}}}{(x_{F}^{\prime})}.
\end {equation}
Here the rescaled quantities are defined as
\begin{equation}
x_{F}^{\prime}=r_{s}x_{F}, \hspace{1cm}
r_{s}=\frac{\sqrt{s}}{\sqrt{s^{\prime}}}.
\end{equation}

Now let us take  account of the energy loss of the color octet
$c\overline{c}$ in the color neutralization process.  If the octet
$c\overline{c}$ pair can lose its energy $\Delta E$, the observed
$J/\psi$  at a given $x_{F}$ comes from a $c\overline{c}$ pair
originally produced at the higher value $x_{F}+\Delta x_{F}$[13],
with
\begin{equation}
\Delta x_{F}=\frac{\Delta E}{E_{p}'},
\end{equation}
where $E_{p}'$ is the incoming proton energy corresponding to the
center of mass system energy $\sqrt{s'}$ which has been modified by
the initial state proton energy loss effect.

Combining above ingredients on initial  and final state effects, the
$J/\psi$ production  cross section in proton-nucleus collisions can
be expressed as
\begin{equation}
\langle \frac{d\sigma}{d x_{_{\rm
F}}}\rangle=\sum\limits_{n=1}^{A}P(n)\frac{d\sigma^{(n)}_{pA}}{d
x_{_{\rm F}}}(x_F).
\end{equation}
Here $P(n)$ is the probability of a projectile proton having $n$
collisions in  nuclei[10,11].

\section{Results and discussion}
The E866 collaboration [3,4] published the precise measurement on
the differential cross-section ratios $R_{W/Be}(x_{F})$,
\begin{equation}
R_{W/Be}(x_{F})=\frac{\langle \frac{d\sigma^{p-W}}{d x_{_{\rm
F}}}\rangle}{\langle \frac{d\sigma^{p-Be}}{d x_{_{\rm F}}}\rangle},
\end{equation}
for proton-induced tungsten to beryllium target for $J/\psi$
production with very broad coverage $-0.1<x_{F}<0.95$. With a proper
color neutralization time $\tau_{0}=0.25$ fm[9], the distance which
the octet $c\overline{c}$ pair travels during this time exceeds the
length of its path through the medium for the kinematical domain
$x_{F}>0$, which implies that the $c\overline{c}$ will remain
colored on its entire path in the medium. The experimental data on
$R_{W/Be}(x_{F})$ will be used to study the  energy loss incurred by
the octet  $c\overline{c}$ pair propagating through the nucleus.

If only considering the nuclear effects of parton distribution
functions, the calculated $J/\psi$ production cross section ratios
$R_{W/Be}(x_{F})$(dotted line)  are compared with the E866
experimental data in Fig.1  by using EPS09 nuclear parton
distribution functions[14] together with CTEQ6L parton density in
the proton[15]. The calculated result shows that the nuclear
suppression  from the nuclear effects on the parton distribution
functions  becomes larger as the increase of $x_{F}$ in the range
$x_{F}>0$.  The nuclear suppression is approximately  from zero  to
$17\%$ for $R_{W/Be}(x_{F})$  in the range  $0 \leq x_{F}\leq 0.95$.

Further, taking into account the energy loss of the beam proton in
initial state,  our calculations use  the center-of-mass system
energy loss of per collision $\triangle\sqrt{s}=$  0.18 GeV
determined from the nuclear Drell-Yan experimental data in the
Glauber model[11]. It is shown that the nuclear suppression on
$R_{W/Be}(x_{F})$ from the energy loss of the beam proton in initial
state increases gradually in the region $0\leq x_{F}\leq 0.8$, and
becomes much steeper in the region $x_{F}>0.8$. The nuclear
suppression due to the beam proton energy loss effect is
approximately $4\%$ to $17\%$  in the range $0\leq x_{F}\leq 0.8$
and $17\%$ to $44\%$  in the range $0.8\leq x_{F}\leq 0.95$. As can
be seen by the dashed line in Fig.1, the total suppression from  the
two kinds of initial state nuclear effects is roughly $4\%$ to
$30\%$ and $30\%$ to $60\%$ in the ranges  $0\leq x_{F}\leq 0.8$ and
$0.8\leq x_{F}\leq 0.95$, respectively.

Next we study the energy loss $\Delta E$ incurred by the octet
$c\overline{c}$ pair propagating through the nucleus. In order to
compare with the energy loss of light quark in cold nuclear matter,
two different parametrizations for  energy loss were proposed with
the help of similarity cases in the nuclear Drell-Yan process
[16,17]. One is written as
\begin{equation}
\Delta E=\alpha L_{A},
\end{equation}
which denotes that the energy loss of the $c\overline{c}$  pair is
liner with the path length $ L_{A}$. Another one is presented as
\begin{equation}
\Delta E=\beta L_{A}^{2}.
\end{equation}
Obviously, the energy loss is quadratic with the  path length. Here,
 $\alpha$ and $\beta$ are the parameters that can be extracted from
experimental data.

In order to determine the  parameter $\alpha$ and $\beta$ in the
energy loss expressions, the calculated $J/\psi$ production
cross-section ratios $R_{W/Be}(x_{F})$ are compared with the
selected experimental values by using the CERN subroutine MINUIT
[18] and minimizing $\chi^2$,
\begin{equation}
\chi^{2}=\sum_{i}^{N} \Bigg[\frac{R^{data}_{W/Be, i}-R^{theo}_{W/Be,
i}}{\sigma^{err}_{i}} \Bigg]^2,
\end{equation}
where $R^{data}_{W/Be, i}$ and $R^{theo}_{W/Be, i}$ indicate
separately the experimental data and theoretical values of the
cross-section ratios $R_{W/Be}$, $\sigma^{err}_{i}$ is the
uncertainty of the experimental points.   Our analysis has in total
27 data points from the E866 experiment in the kinematical range
$x_{F}>0$. One standard deviation of the optimum parameter
corresponds to an increase of $\chi^{2}$ by 1 unit from its minimum
$\chi^{2}_{min}$. The fit of the selected data makes
$\alpha=3.42\pm0.15$  with the relative uncertainty
$\delta\alpha/\alpha\simeq 4\%$ and $\chi^{2}/ndf=20.56$. Regarding
the quadratic energy loss expression of the $c\overline{c}$ pair,
the value of  parameter $\beta$ as well as its corresponding error
extracted from the fit of the experimental data is
$\beta=0.55\pm0.03$ with the relative uncertainty
$\delta\alpha/\alpha\simeq 5\%$ and $\chi^{2}/ndf=20.52$.

The solid line in Fig.1 shows  the  calculated $R_{W/Be}(x_{F})$ for
$J/\psi$ production, compared with the E866 data, given by using the
EPS09  nuclear parton distributions together with the energy loss of
the beam proton in initial state  and the linear color octet
$c\overline{c}$ energy loss. The theoretical value of the  cross
section on $R_{W/Be}(x_{F})$ from the linear  energy loss has almost
no difference with that given by the quadratic energy loss. The
total nuclear suppression on $R_{W/Be}(x_{F})$ from the three
nuclear effects increases gradually from $11\%$ to $60\%$ in the
region $0\leq x_{F}\leq 0.8$, and gets rapidly bigger from $60\%$ to
$96\%$ in the region $0.8\leq x_{F}\leq 0.95$.

\begin{figure}[!]
\centering
\includegraphics*[width=11cm, height=10cm]{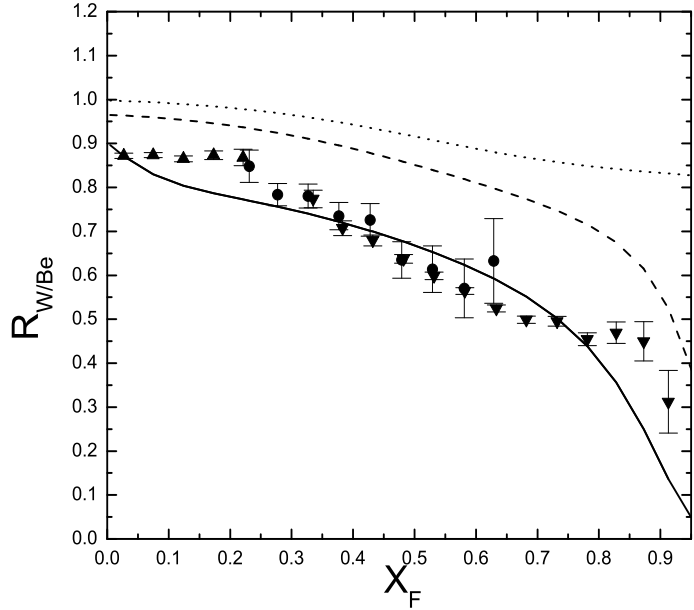}
\vspace{-0.5cm} \caption{The calculated $J/\psi$ production
cross-section ratios $R_{W/Be}(x_{F})$ with three nuclear
effects(solid line), the nuclear effects on the parton distribution
functions and energy loss of beam proton ( dashed line), and only
the nuclear effects on the parton distribution functions (dotted
line). The solid triangles ( circles and inverted triangles) are the
E866 experimental data[3,4] in  the region $0<x_F < 0.3$  ( $0.2
<x_F < 0.65$ and $0.3 <x_F < 0.95$).}
\end{figure}

However, whether the linear color octet $c\overline{c}$ energy loss
or the quadratic energy loss, the value of $\chi^{2}/ndf$ extracted
from the selected data on $R_{W/Be}$ is too high in spite of the
small  relative uncertainty. From the Fig.1, we can find intuitively
that the numerical calculations of the cross-section ratios
$R_{W/Be}(x_{F})$ deviate far from the experimental data in small
($x_{F}<0.2$) and large ($x_{F}>0.8$) $x_{F}$ regions. For the
quantitative evaluation of the possible origins for the big
$\chi^{2}/ndf$ value, the $\chi^{2}$ analysis on the experimental
data is performed by means of the different combination of three
$x_{F}$ regions for the small($0<x_F < 0.3$), middle ( $0.2 <x_F <
0.65$) and big ($0.3 <x_F < 0.95$) $x_{F}$ regions. The values of
parameters $\alpha$, $\beta$   and $\chi^2$ per degrees of freedom
are calculated and summarized in Table I by fitting the selected
experimental data on $R_{W/Be}$ as a function of $x_{F}$. As can be
seen from the Table I, the big $\chi^{2}/ndf$ value really stems
from the experimental measurement in  the small($0<x_F < 0.3$) and
big ($0.3 <x_F < 0.95$) $x_{F}$ regions. If only using the
experimental data in middle $x_{F}$ regions, the  $\chi^{2}/ndf$
value is approximately equal to one for the linear and quadratic
color octet $c\overline{c}$ energy loss. The obtained
$\alpha=2.78\pm0.81$  with the relative uncertainty
$\delta\alpha/\alpha\simeq 29\%$, and $\beta=0.44\pm0.13$  with
$\delta\beta/\beta\simeq 29\%$.

\begin{table}[t,m,b]
\caption{The values of parameters $\alpha$, $\beta$   and
$\chi^{2}/ndf$ extracted from the selected data on $R_{W/Be}$ as a
function of  $x_{F}$.}
\begin{center}
\begin{tabular}[c]{p{4cm}p{3cm}p{4cm}p{4cm}c}\hline  \hline

          $x_{F}$ &No.data& $\alpha(\chi^{2}/ndf)$ &$\beta(\chi^{2}/ndf)$\\
         \hline
       {0.20-0.65}& 9 & $2.78\pm0.81(1.01)$&$0.44\pm0.13$(1.03)\\
   {0.30-0.95}&13 &$4.44\pm0.21(13.29)$&$0.71\pm0.03$(12.32)\\
       {0.20-0.95} &22 & $4.35\pm0.20$(8.78)&$0.70\pm0.03$(8.35)\\
{0.00-0.95}&27 & $3.42\pm0.15(20.56)$&$0.55\pm0.03(20.52)$\\

\hline  \hline
\end{tabular}
\end{center}
\end{table}

Regarding the energy loss of a charm quark, if it is assumed that
the charm and anti-charm quark lose the same energy, the obtained
energy loss per unit length is $\alpha=1.39$ GeV/fm. For an outgoing
light quark, however, the theoretical energy loss  is $0.38$ GeV/fm
 by the global fit of the experimental data in semi-inclusive deep
inelastic scattering of leptons on the nucleus[19]. It is indicated
that the heavy quark in cold nuclear matter loses more energy
compared to the light quark.

\begin{figure}[!]
\centering
\includegraphics*[width=11cm, height=10cm]{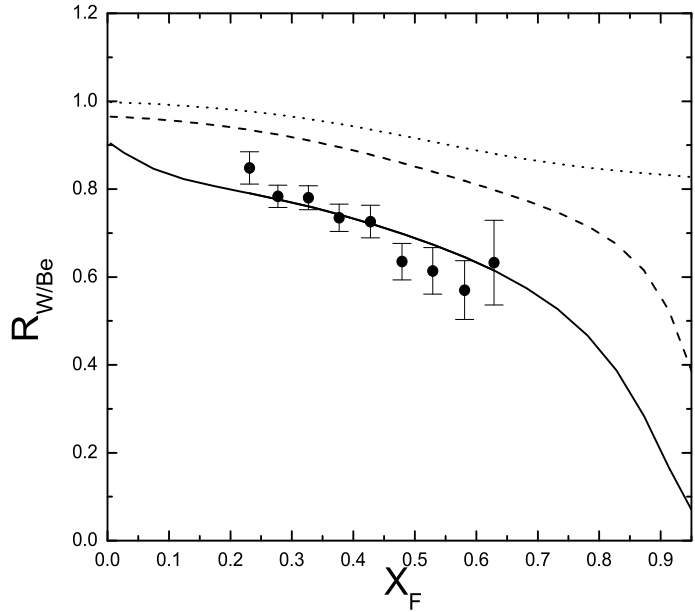}
\vspace{-0.5cm} \caption{ The $J/\psi$ production cross-section
ratio $R_{W/Be}(x_{F})$.  The filled circles are the E866
experimental data [3,4] in the region $0.2 <x_F < 0.65$. The
comments are the same as Fig.1. }
\end{figure}

By making use of the linear  color octet $c\overline{c}$ energy loss
determined from the E866 experimental data in the region $0.2 <x_F <
0.65$, we revisit the final state energy loss effect on the $J/\psi$
production cross-section ratio $R_{W/Be}(x_{F})$. The theoretical
results are compared with the relative experimental data  in Fig.2.
It is shown that the nuclear effects in the initial state provide
roughly the $J/\psi$ suppression from  $4\%$ to $15\%$, $15\%$ to
$60\%$ in the ranges  $0< x_{F}< 0.5$ and $ 0.5< x_{F}< 0.95$,
respectively. The contributions from nuclear effects in the final
state is approximately from $6\%$ to $15\%$ and $15\%$ to $34\%$  in
the corresponding $x_{F}$ ranges. Therefore, it is demonstrated that
the $J/\psi$ suppression on $R_{W/Be}(x_{F})$ induced by the energy
loss of color octet $c\overline{c}$ is more important than that of
the initial state nuclear effects in the $x_{F}<0.5$ region, but the
exact reverse is the case in the $ x_{F}> 0.5$ region.

\section{ Summary }

$J/\psi$ production in p-A collisions is an ideal tool for exploring
the energy loss of charm quarks in cold nuclear matter. In this
paper, the nuclear effects for $J/\psi$ production in p-A collisions
are investigated by considering the nuclear effects on parton
distribution, energy loss of beam proton and the finial state energy
loss of color octet $c\overline{c}$ pair, respectively.  We perform
a leading order phenomenological analysis on the $J/\psi$ production
cross-section ratio $R_{W/Be}(x_{F})$ and compare with the E866
experimental data for the $c\overline{c}$ remaining colored on its
entire path in the nuclear  medium. It is found that the theoretical
results by combining the different nuclear effects has a good
agreement with the experimental data. The present experimental data
do not distinguish between the linear and quadratic dependence of
$c\overline{c}$ pair energy loss. The $J/\psi$ suppression on
$R_{W/Be}(x_{F})$ from the initial state nuclear effects is more
important than that induced by the energy loss of color octet
$c\overline{c}$ in the large $x_F$ region. By means of the selected
experimental data, we obtain the $c\overline{c}$ pair energy loss
per unit path length $\alpha=2.78\pm0.81$ GeV/fm, which indicates
that the heavy quark can lose more energy compared to the outgoing
light quark in cold nuclear matter.  We desire that our research
results  can provide useful reference for deep understanding of
$J/\psi$ suppression in heavy-ion collisions.

\vskip 0.5cm
{\bf Acknowledgments}
This work was supported in part by the National Natural Science
Foundation of China(11075044, 11347107)and Natural Science
Foundation of Hebei Province (A2013209299).

\end{document}